\documentclass[english,prb,twocolumn,superscriptaddress,groupedaddress]{revtex4}
\usepackage[utf8]{inputenc} 
\usepackage[T1]{fontenc}    
\usepackage{url}            
\usepackage[colorinlistoftodos, color=green!40, prependcaption]{todonotes}
\usepackage{color}
\usepackage{lmodern}
\usepackage{babel}
\usepackage{float}
\usepackage{graphicx}
\usepackage{xcolor}
\usepackage{amsmath}
\usepackage{amssymb}
\usepackage{gensymb}
\usepackage{mathtools}
\definecolor{burgundy}{rgb}{0.5, 0.0, 0.13}
\usepackage[linktoc=page,colorlinks=true,allcolors=burgundy]{hyperref}
\makeatletter

\begin{document}
\title{Electronic transport in sub-micrometric channels at the LaAlO$_{3}$/SrTiO$_{3}$ interface}

\author{Margherita Boselli\textsuperscript{{*}}}
\address{Departement of Quantum Matter Physics, University of Geneva, 24 Quai
Ernest-Ansermet, 1211 Geneva, Switzerland}
\author{Gernot Scheerer}
\address{Departement of Quantum Matter Physics, University of Geneva, 24 Quai
Ernest-Ansermet, 1211 Geneva, Switzerland}
\author{Michele Filippone}
\address{Departement of Quantum Matter Physics, University of Geneva, 24 Quai
Ernest-Ansermet, 1211 Geneva, Switzerland}
\author{ Weiwei Luo}
\address{Departement of Quantum Matter Physics, University of Geneva, 24 Quai
Ernest-Ansermet, 1211 Geneva, Switzerland}
\address{Now at Nankai University,Tianjin, China}
\author{Adrien Waelchli}
\address{Departement of Quantum Matter Physics, University of Geneva, 24 Quai
Ernest-Ansermet, 1211 Geneva, Switzerland}
\author{ Alexey B. Kuzmenko}
\address{Departement of Quantum Matter Physics, University of Geneva, 24 Quai
Ernest-Ansermet, 1211 Geneva, Switzerland}
\author{Stefano Gariglio}
\address{Departement of Quantum Matter Physics, University of Geneva, 24 Quai
Ernest-Ansermet, 1211 Geneva, Switzerland}
\author{Thierry Giamarchi}
\address{Departement of Quantum Matter Physics, University of Geneva, 24 Quai
Ernest-Ansermet, 1211 Geneva, Switzerland}
\author{Jean-Marc Triscone}
\address{Departement of Quantum Matter Physics, University of Geneva, 24 Quai
Ernest-Ansermet, 1211 Geneva, Switzerland}

\email{margherita.boselli@unige.ch}

\begin{abstract}
Nanoscale channels realized at the conducting interface between LaAlO$_{3}$ and SrTiO$_{3}$ provide a perfect playground to explore the effect of dimensionality on the electronic properties of complex oxides. Here we compare the electric transport properties of devices realized using the AFM-writing technique and conventional photo-lithography. We find that the lateral size of the conducting paths has a strong effect on their transport behavior at low temperature. We observe a crossover from  metallic to  insulating regime occurring at about 50~K for channels narrower than 100~nm. The insulating upturn can be suppressed by the application of a positive backgate. 
We compare the behavior of nanometric constrictions in lithographically patterned channels with the result of model calculations and we conclude that the experimental observations are compatible with the physics of a quantum point contact. 
\color{black} 
\end{abstract}

\maketitle

\section{Introduction}
At oxide interfaces, the interplay between confinement effects and the physical properties emerging from the junction of the constituent materials gives rise to a plethora of phenomena unattainable in III-V semiconductor heterostructures. One of the most studied systems is the conducting two dimensional electron system (2DES) appearing at the interface between LaAlO$_{3}$ and SrTiO$_{3}$ \cite{ohtomo_2004}. It exhibits a series of interesting properties including gate-tunable superconductivity and spin-orbit coupling \cite{reyren_superconducting_2007, caviglia_electric_2008, caviglia_tunable_2010}. Moreover, recent spin-to-charge conversion experiments, performed both in  LaAlO$_{3}$/SrTiO$_{3}$ heterostructures and at the reduced surface of SrTiO$_{3}$, evidence the potential of this interface for spintronics and topology studies \cite{lesne_highly_2016, chauleau_2016, song_observation_2017, vaz_topology_2019, trier_electric-field_2020}. 
Fundamentals to several studies and application is the capapability to reduce the lateral dimension of the conducting channels. \\
Among the different methods used to nanostructure the 2DES, which include photo-lithography \cite{rakhmilevitch_phase_2010}, and electron-beam lithography \cite{stornaiuolo_plane_2012, aurino_nano-patterning_2013, goswami_quantum_2016}, 
the so-called “AFM-writing” technique offers a versatile way to sketch conducting nano-devices in LaAlO$_{3}$/SrTiO$_{3}$ heterostructures \cite{cen_nanoscale_2008, cen_oxide_2009}. It relies on the application of a positive voltage to the tip of an atomic force microscope (AFM), while scanning the LaAlO$_{3}$ surface, to realize conducting patterns at the otherwise insulating interface between SrTiO$_{3}$ and 3 unit cells (u.c.) of LaAlO$_{3}$.\\ These nano-lithographic techniques give access to the study of interesting phenomena including universal conductance fluctuations \cite{rakhmilevitch_phase_2010,stornaiuolo_plane_2012}, the Josephson effect in planar junctions \cite{goswami_nanoscale_2015, goswami_quantum_2016, monteiro_side_2017, stornaiuolo_2017}, superconductivity in AFM-written nanowires \cite{veazey_superconducting_2012,veazey_oxide-based_2013}  and conductance quantization in quantum-point contacts (QPCs) \cite{ron_2014, tomczyk_2016, thierschmann_2018, jouan_QPC_2020}. 
Among the different phenomena recorded in these nano-devices, there are a few effects that deserve further investigation, for instance: in AFM-written nanowires it has been suggested the presence of electron pairs without superconductivity \cite{cheng_electron_2015}, and a surprisingly long ballistic length has been observed in the normal state of nano-channels \cite{ron_2014, tomczyk_2016}. Furthermore, the role of the lithography technique on the channel properties has to be clarified, since nanowires realized via electron-beam lithography often display an insulating state at low temperature  \cite{aurino_nano-patterning_2013, aurino_retention_2016}. \\
We report here on a comparative study of electric transport in (sub-)micrometric conducting channels realized at the LaAlO$_{3}$/SrTiO$_{3}$  interface using two methods: the AFM-writing technique and conventional photo-lithography. The lateral size of AFM-written wires ranges from 50 to 200~nm \cite{boselli_characterization_2016} and is comparable with the characteristic electronic length-scales of the 2DES at low temperature. Indeed, the inelastic scattering
length is $\sim$200~nm at 1~K \cite{rakhmilevitch_phase_2010, stornaiuolo_plane_2012}, the elastic mean-free path ranges from 10 to 100~nm at 1.5~K \cite{stornaiuolo_weak_2014}, and the superconducting coherence length  from 10 to 70~nm \cite{reyren_superconducting_2007, li_probing_2018}. The temperature behavior of the channel resistance displays a clear size-dependence: wires narrower than $\sim$100~nm have a crossover from a metallic to an insulating behavior at approximately 50~K, while wider wires stay metallic down to 1.5~K. \\ To shed light on this effect, we also investigate devices realized  by conventional photo-lithography. These paths have a minimal width of $\sim$1 $\mu$m,  i.e. larger than those realized via AFM-writing, but their conductance is more stable in time. One of these devices is found to have a few $\sim$50~nm wide bottle-neck constrictions giving rise to an insulating behavior below 45~K. Moreover, at 50~mK, field effect experiments revealed that the electronic transport is characterized by conductance quantization through these constrictions. \\ By comparing these experimental results with calculations for a quantum point contact (QPC) with a saddle-point potential,  we show that the conductance quantization originates from the lateral confinement of the electrons at the interface; such lateral confinement can also explain the transport properties below 50~K. 
We conclude that constrictions acting as QPCs, similar to those reported in lithographic devices, could also be at the origin of the insulating behavior observed in AFM-written nanowires.\\  
The paper is structured as follows:  the experimental methods are described in Section II and Section III presents the experimental results. Section IV describes the theoretical model developed to interpret the data, and finally we will discuss our findings in Section V.\\

\section{Experimental details}
The LaAlO$_{3}$ thin films have been grown by pulsed laser deposition (PLD)  on commercial TiO$_{2}$-terminated SrTiO$_{3}$  substrates provided by Crystec GmbH, using a laser fluence of 0.6 J/cm$^{2}$ and a repetition rate of 1~Hz. During the deposition, the substrates are kept at 800~$^{\circ}$C in an O$_{2}$ pressure of 10$^{-4}$~mbar. After the deposition, the samples are annealed \textit{in-situ} in an O$_{2}$ pressure of 200~mbar at 550~$^{\circ}$C for 1 hour, and later cooled down to room temperature in the same atmosphere. The LaAlO$_{3}$ thickness, crucial for the AFM-writing technique, is monitored \textit{in-situ} using the reflection high-energy electron diffraction (RHEED) technique.\\   
After the growth, AFM-written channels are prepared (see for details reference \cite{boselli_characterization_2016}) at room temperature by scanning the AFM tip between conducting electrodes in contact with the interface. The channels' width is controlled by the voltage applied to the AFM tip (5-9~V,) and by the number of consecutive scans, and 
estimated at room temperature using the so-called cutting method. They are typically 10~$\mu$m long and  50~nm-200~nm wide.  All the transport measurements are performed using a four-point configuration as shown in Figure \ref{fig:fig1}(a). \\
Devices realized with photo-lithography were prepared using the following procedure: first a layer of photoresist, lithographically patterned with the shape of the conducting device, covers a bare substrate, then amorphous SrTiO$_{3}$ is deposited by PLD at room temperature in an O$_{2}$ pressure of 10$^{-4}$~mbar, finally the photoresist is removed to expose the selected areas of the  substrate to the deposition of crystalline LaAlO$_{3}$. 
The structures realized with this technique are Hall-bars  $\sim$1~$\mu$m wide and  up to 300~$\mu$m long (see Figure \ref{fig:fig1}(d)). While their lateral size is in general larger than the characteristic electronic length-scales of the 2DES, their length/width ratio is chosen to mimic that of AFM-written wires. \\ 
For field effect tuning, we sputter  a few $\mu$m of gold on the backside of the SrTiO$_{3}$ substrate to create a gate electrode.

\section{Results}
\begin{figure*}
 \begin{center}
  \includegraphics[width=1.5\columnwidth]{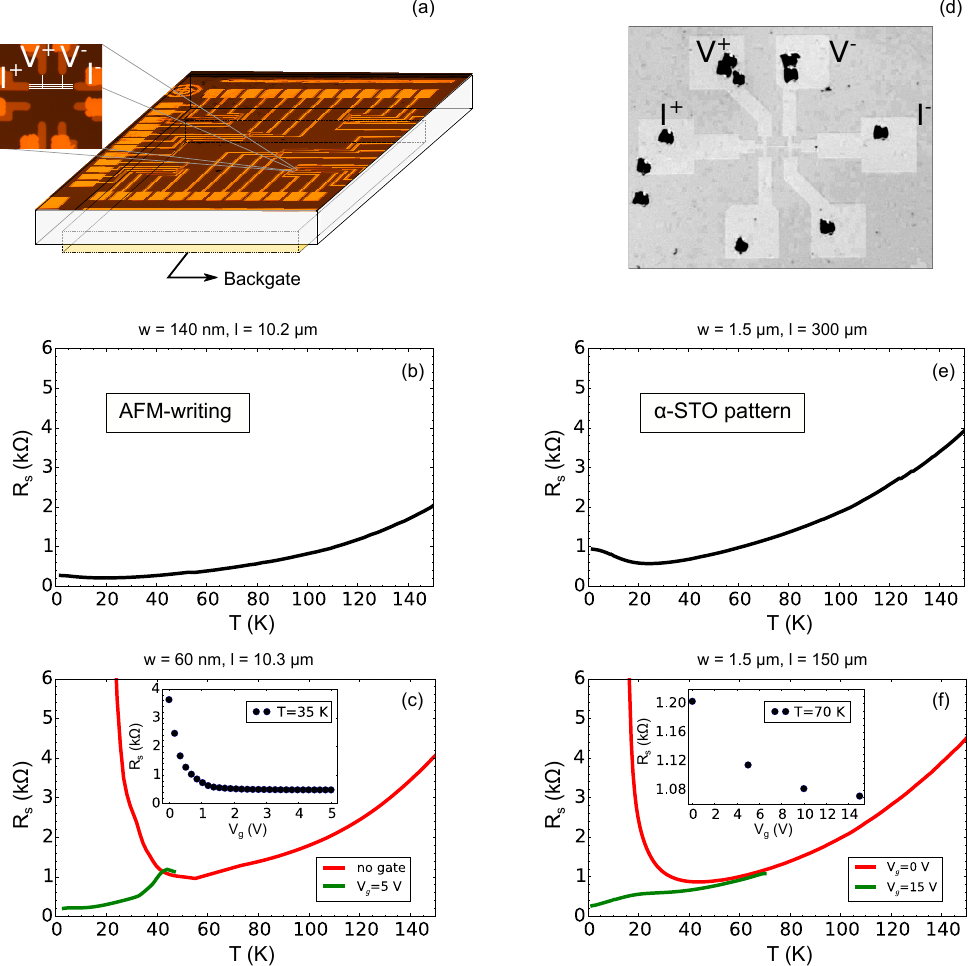} \\
\caption{\label{fig:fig1} (a) Sketch of an AFM-written device. A zoom of the writing area surrounded by metallic electrodes is shown in the inset. The sheet resistance as a function of temperature between 150~K and 1.5~K for AFM-written wires of different widths is shown in panels (b) and (c). (b) Two parallel AFM-written wires 10.2~$\mu$m long, and 140~nm wide and 1~$\mu$m apart from each other. (c) Three parallel AFM-written wires 10.3~$\mu$m long, 60~nm wide, and 1~$\mu$m apart from each other, with (green curve) and without (red curve) an applied gate voltage. The inset shows the evolution of the sheet resistance upon backgate voltage at 35~K. We note that the sheet resistance at 35~K and $V_{g}$=0 is $\sim$4~k$\mathrm{\Omega}$, hence higher than $R_{s}$ at 35~K extracted from the red curve. This discrepancy is related to the thermal history of the device: first it was cooled down to 20~K (red curve), and then warmed up again to 35~K, where $R_{s}$ was found to be higher than at the beginning of the process. 
(d) Optical image of a Hall-bar realized with a pattern of amorphous SrTiO$_{3}$.  The channel is nominally 1.5~$\mu$m wide and 150~$\mu$m long. Panels (e) and (f) show the sheet resistance as a function of temperature between 150~K and 1.5~K of two lithographically-patterned devices with different sizes. (e) R$_{\mathrm{s}}$(T) of a channel 300~$\mu$m long, and 1.5~$\mu$m wide. (f) R$_{\mathrm{s}}$(T) of a device 150~$\mu$m long, 1.5~$\mu$m wide and containing some constrictions formed by insulating islands in the middle of the conducting region (see Figure \ref{fig:fig2}). The evolution of the sheet resistance as a function of the gate-voltage at 70~K is shown in the inset.}
 \end{center}
\end{figure*}

\begin{figure*}
 \begin{center}
  \includegraphics[width=1.8\columnwidth]{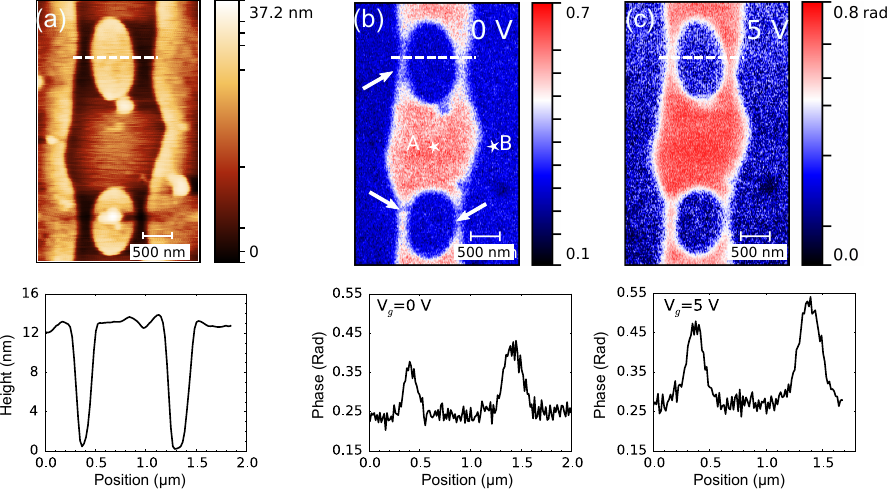} \\
 \end{center}
\caption{\label{fig:fig2} (a) AFM topography of the conducting channel realized with a pattern of amorphous SrTiO$_{3}$ in correspondence to the constrictions. (b) Phase component of the near-field optical signal measured using a cryo-SNOM at 5~K, with an incident light wavelength of 10.7~$\mu$m, and the backgate grounded. The insulating regions at the center of the constrictions are indicated by white arrows. The symbols labelled A and B indicates the positions considered to evaluate the phase difference between the wide conducting channel and the insulating region. (c) Phase component of the near-field optical signal with 5~V applied to the backgate. The plots below panels (a), (b) and (c) show a line cut of the topography and the phase component of the near-field signal at V$_{g}$=0~ and 5~V, respectively. The dashed white lines in the main Figures indicate the position of the cuts.}
\end{figure*}

\begin{figure*}
 \begin{center}
  \includegraphics[width=1.5\columnwidth]{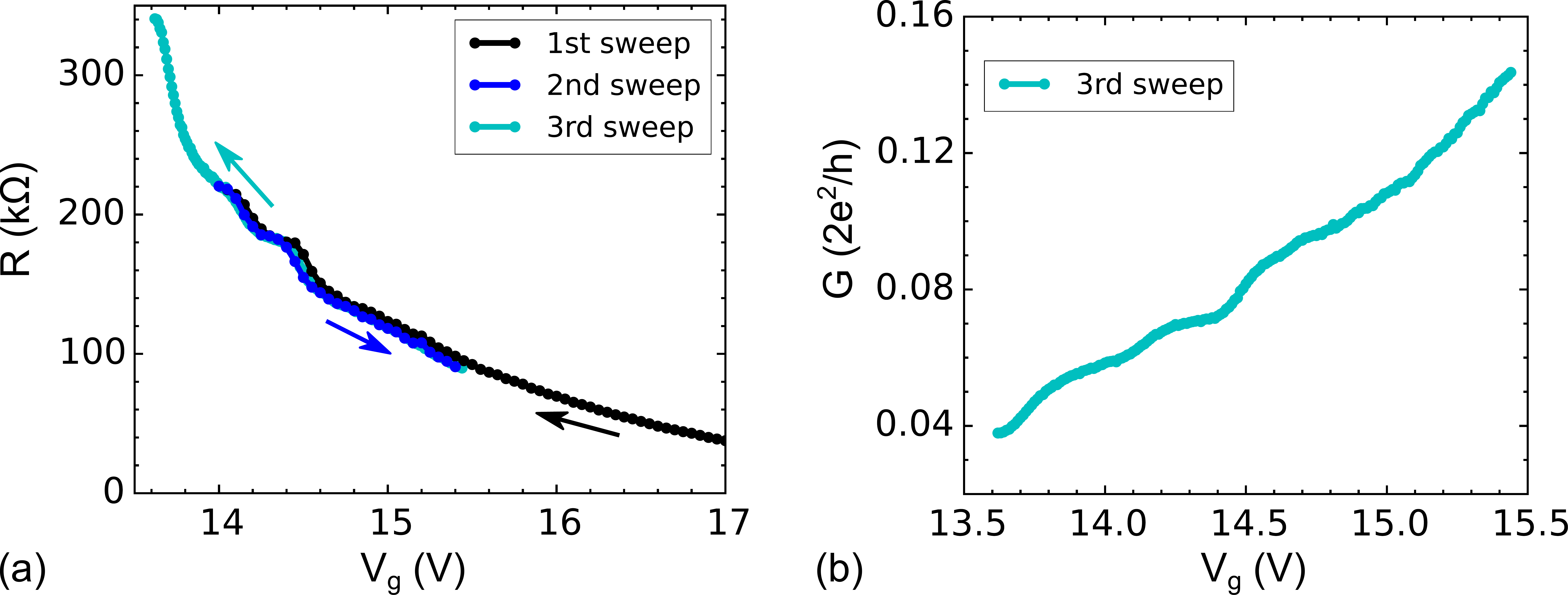} \\
 \end{center}
\caption{\label{fig:fig3}(a) Backgate dependence of the total resistance at 50~mK for the device realized with a pattern of SrTiO$_{3}$. Three datasets have been acquired successively by sweeping the backgate voltage up and down (the arrows indicate the sweeping direction). (b) Gate dependence of the total conductance (for the third sweep only), normalized to the quantum of conductance 2$e^{2}/h$.}
\end{figure*}

\begin{figure*}
 \begin{center}
  \includegraphics[width=1.5\columnwidth]{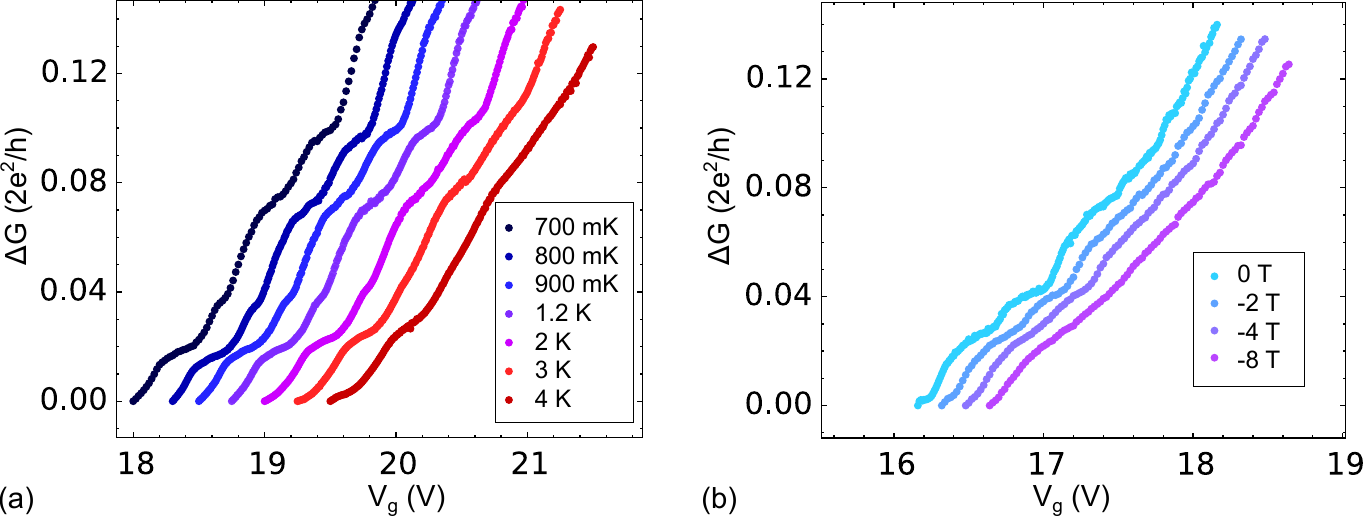} \\
 \end{center}
\caption{\label{fig:fig4} Backgate dependence of the conductance increase $\Delta G= G(V_{g})- G(V^{min}_{g})$ normalized to 2$e^{2}/h$ acquired at increasing temperatures (a), and increasing magnetic fields at 400~mK (b). The curves have been shifted along the x-axis for clarity.}
\end{figure*}

Figure \ref{fig:fig1} shows the temperature dependence of the sheet resistance of four devices fabricated using the AFM-writing technique (b,c) or
patterns of amorphous SrTiO$_{3}$ (e,f). 
The behavior of the AFM-written wires depends on their channel width.  Wires wider than $\sim$100~nm have an overall metallic behavior (see Panel \ref{fig:fig1}(b)), while narrow channels  have a crossover to an insulating behavior
occurring at approximately 50~K (see Panel \ref{fig:fig1}(c)). This transition can be suppressed (metallicity kept down to 1.5~K) by the application of a positive voltage to the backgate (see Figure \ref{fig:fig1}(c)). All the devices realized with patterns of amorphous SrTiO$_{3}$ are metallic with a finite resistance at low temperature (see Figure \ref{fig:fig1}(e)), except for one:  a device nominally 1.5~$\mu$m wide and 150~$\mu$m long has a crossover from a metallic to an insulating behavior at $\sim$50~K.
\color{black} A positive gate voltage restores the metallicity (see Figure \ref{fig:fig1}(f)).\\ To understand this surprising behavior in a lithographically defined channel, similar to that of narrow AFM-written wires,
we investigated the topography of this device and revealed the presence of two consecutive oval structures located close to one end of the path (see  Figure \ref{fig:fig2}(a)). These regions have exactly the same height as the regions covered by the amorphous SrTiO$_{3}$ used to prevent the formation of the 2DES at the interface (inset of Figure \ref{fig:fig2}(a)). We believe that these islands originate from residues of photoresist left on the substrate, and we deduce that they are insulating regions creating a series of conducting constrictions 0.5-1.5~$\mu$m long and from 50 to 500~nm wide in the otherwise wider conducting channel.\\ The 2DES conductivity in proximity to these features has been investigated using the scattering-type scanning near-field optical microscopy (s-SNOM) technique \cite{luo_high_2019}. Figures \ref{fig:fig2}(b) and (c) show two local maps, measured at 5~K, of the phase component of the s-SNOM signal. In the energy range probed here (laser wavelength of 10.7~$\mu$m), the phase component of the optical response is a reliable measure of the interface local conductivity \cite{luo_high_2019}. This analysis shows that the conductivity of the  constrictions cooled down with the gate grounded is much smaller than that of the larger area of the channel.
Indeed, the s-SNOM phase difference between the center of the bottle-neck and the insulating regions around the channel, $\sim 0.14$~rad, is low compared to the difference referred to the wide conducting channel, $\sim 0.32$~rad. 
By increasing the backgate voltage to 5~V, the phase difference between the constrictions and the insulating regions increases to $\sim 0.2$~rad, (cf. intensity of the peaks in the horizontal cuts at the bottom of Figure \ref{fig:fig2} at $V_{g}$=0~V and $V_{g}$=5~V), while the same quantity estimated at the center of the large channel raises to $\sim 0.38$~rad,
indicating that the constrictions become more metallic. These observations suggest that the insulating behavior of this device originates from the regions where the channel width is reduced. \\
Taking advantage of the nanosize channels created by the bottle-neck structures of amorphous SrTiO$_{3}$, we measured the electronic transport below  1~K in a dilution cryostat. The sample was cooled down to 50~mK with  14~V applied to the backgate and all the measurements were performed with a constant current of 8~nA applied between the drain and the source. We note that this sample does not display any sign of superconductivity in the set-up used for the measurements. This may be related to the set-up (i.e. lack of appropriate filters or very low critical current) or to an intrinsic reason such as doping, low transition temperature or 1D behavior. 
Figure \ref{fig:fig3}(a) shows  the resistance  as a function of  gate voltage at 50~mK while decreasing $V_{g}$ from $17~\mathrm{V}$ to $13.6~\mathrm{V}$.  The resistance increases from $38~\mathrm{k \Omega}$ to $340~\mathrm{k \Omega}$, and, between $V_{g}=15~\mathrm{V}$ and $V_{g}=13.6~\mathrm{V}$, it is characterized by  some step-like features which are not present in wider LaAlO$_{3}$/SrTiO$_{3}$ paths \cite{caviglia_electric_2008}. Such a large resistance increase (by one order of magnitude)  upon a gate change of a few Volts is due to the strong focusing of the electric field applied between the large gate electrode and the narrow ($1.5~\mu$m wide) channel \cite{rakhmilevitch_anomalous_2013}. We assume that the channel resistance results from the sum of  two contributions. One comes from the resistance of  the $1.5~\mathrm{\mu m}$  wide and almost 150~$\mu$m long channel, and the other from  the small constrictions in series with it. We attribute the continuous increase of the resistance to the charge depletion in the full channel and the step-like features to the constrictions. This behavior is 
reproducible upon several gate voltage sweeps and resembles that of the conductance of a QPC upon gate tuning \cite{vanwees_1988}. For a more detailed analysis, we convert the resistance data in units of quantum of conductance, $G$ ($2e^{2}/h$) (see Figure \ref{fig:fig3}(b)). $G$ shows three steps  approximately $0.02\cdot(2e^{2}/h)$  high and separated by roughly ${\Delta V_{g}}=350~\mathrm{mV}$. 
The height of the steps is much smaller than  $2e^{2}/h$, and this discrepancy is attributed to the contribution of the full channel to the total conductance (the $1.5~\mu m$ wide and almost 150~$\mu$m long conduction path in series with the constrictions). Therefore, a quantitative comparison with theory requires further analysis as presented in the next section.  \\       
Figure \ref{fig:fig4}(a) shows the constriction conductance as a function of the gate voltage at different temperatures, where four steps in $G(V_{g})$ are visible. These datasets have been acquired a few days after the one shown in \ref{fig:fig3}(b), and the measurements are characterized by smaller noise levels and a slightly different step shape. The conductance steps survive up to 1~K, they are smeared out at higher temperature, and eventually vanish at 4~K. The suppression of the conductance quantization in temperature is expected for a QPC \cite{mesoscphys, semicondnanos}, but, interestingly, our device is particularly robust to the effect of temperature.
The application of a magnetic field (oriented out of the interface plane) induces a similar effect: the conductance steps are smeared out at 8~T as shown in Figure \ref{fig:fig4}(b). \\

\section{Numerical model}
\begin{figure*}
\includegraphics[width=1\columnwidth]{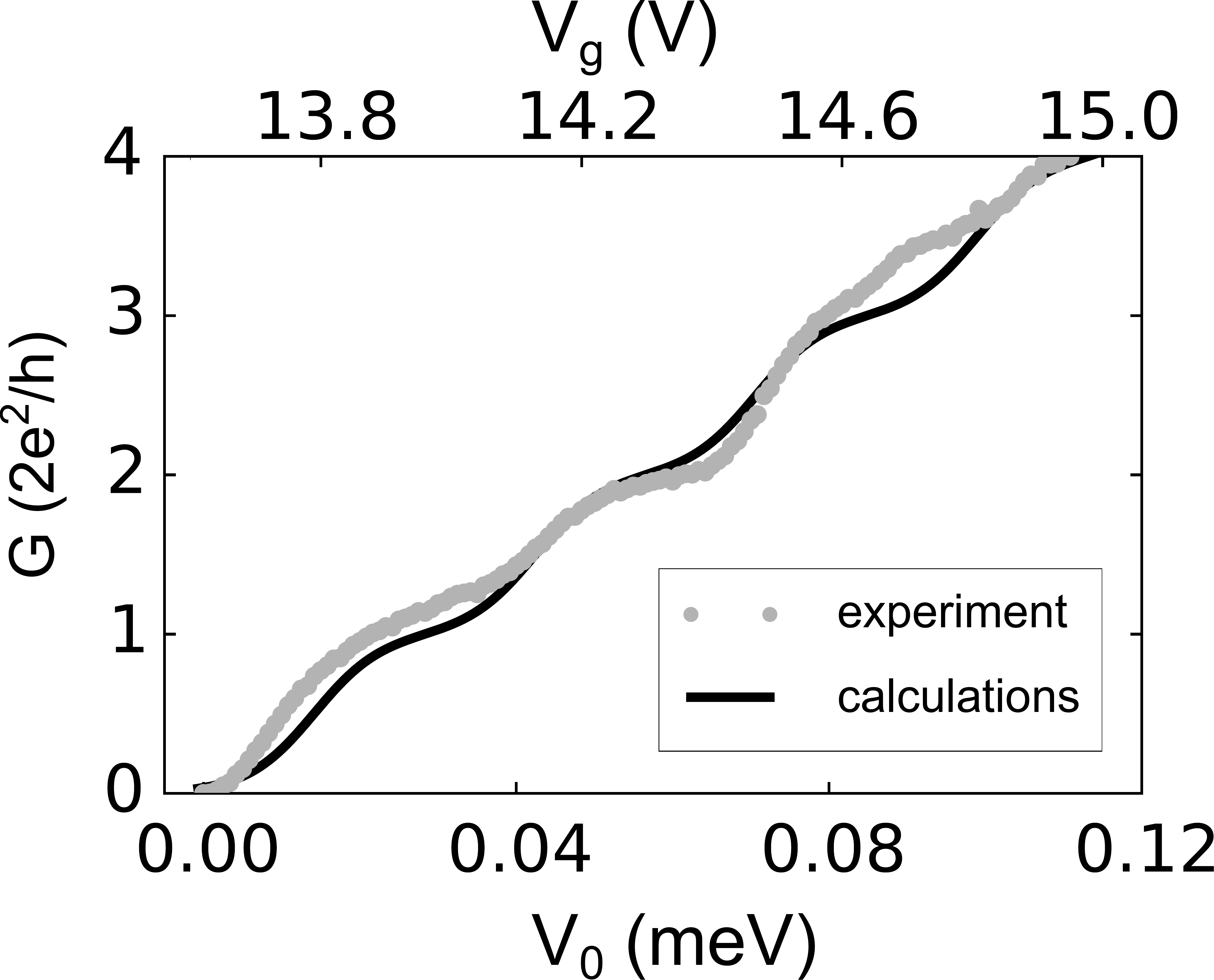} \\
 \caption{\label{fig:fig6} Calculated and experimental values of the conductance of the QPC (in unit of 2$\mathrm{e^{2}/h}$) as a function of $V_{0}$  at 50~mK. The numerical calculations were carried out using the saddle-point model for a QPC, with $l_{x}$=350~nm and $l_{y}$=35~nm. The experimental conductance has been  analysed as explained in the main text. The gate voltage applied to the device is reported on the top ascissa axis for comparison.}
\end{figure*}

We interpret the behavior of the structural constrictions below 1~K in the framework of the saddle-point model for a QPC \cite{buttiker_quantized_1990}.
The confining potential used in this approach provides a good approximation of a bottle-neck constriction as found in our lithographically-patterned device  
(see Figure \ref{fig:fig2}(a)). It is a smooth function of the interface plane coordinates, $x$ and $y$ ($x$ is the current direction and $y$ the transverse one), written as: 
\begin{equation}\label{eq:SP}
V(x,y)=V_{0}-\frac{1}{2}m\omega_{x}^{2}x^{2}+\frac{1}{2}m\omega_{y}^{2}y^{2};
\end{equation}
where $V_{0}$ is the electrostatic potential at the saddle-point,  $\omega_{x}$ and $\omega_{y}$  are two frequencies related to the size of the constriction, and $m$ the electron mass. $V(x,y)$ is the potential of a 2D harmonic oscillator along the $y$ axis, and a potential barrier with a parabolic shape along the  $x$ axis, and $\omega_{x,y}$ can be expressed in term of the length $l_{x}$ and the width $l_{y}$ of the constriction: $ \omega_{x,y}=\frac{\hbar}{ml^{2}_{x,y}}$. 
By solving the eigenvalue problem relative to equation  \eqref{eq:SP}, one can compute the total conductance at finite temperature of a QPC using the Landauer formula \cite{fisher_relation_1981}: 
\begin{equation}\label{eq:cond}
 G=\frac{2e^{2}}{h}\sum_{n}\int d\epsilon \frac{\partial f_{\mu}(\epsilon)}{\partial \epsilon}t_{n}(\epsilon),
\end{equation}
where $f_{\mu}(\epsilon)=(e^{\beta(\epsilon-\mu)}+1)^{-1}$ is the Fermi distribution in the reservoirs,  $t_{n}(\epsilon)$ the transmission coefficient of the n$\mathrm{^{th}}$ channel at energy $\epsilon$ (one should note that channel intermixing does not occur in this case as the confining potential is quadratic), $\mu$ the chemical potential and $\beta=1/k_{B}T$.\\
We model the experimental behavior of our device by making the assumption that the conductance jumps visible in Figure \ref{fig:fig3}(b) originate from a single constriction. Equation \eqref{eq:cond} has been computed numerically, by fixing a finite  number of channels entering in the sum \footnote{We fixed the number of channels to 100. This value is large enough to describe the  conductance steps observed experimentally (Figure \ref{fig:fig3}(b)), involving the conduction channels at low energy.  Considering a larger number of channels would imply making strong assumptions on the high energy structure of the constriction, which clearly goes beyond the regime of validity of our simple description.}, 
using an electron effective mass of $2.2\cdot m$ \cite{fete_rashba_2012}, and setting  the chemical potential,  $\mu$, to zero. The position of $\mu$ is arbitrary as it is compensated by the value of the electrostatic potential $V_{0}$.
We set the values of $l_{y}$  and $l_{x}$ to 35~nm and 350~nm, respectively. Despite the lack of a precise control over the geometry of our device, these values are compatible with the s-SNOM and topographic images shown in Figure \ref{fig:fig2}.
The $V_{0}$-dependence of the calculated conductance  at 50~mK is  shown in Figure \ref{fig:fig6}. 
The energy difference, $\Delta V_{0}$, between the conductance plateaus reflects the energy spacing between the transverse  states, $\hbar \omega_{y} \propto l_{y}^{-2}$,  the sharpness of the steps is controlled by the temperature and $\hbar \omega_{x} \propto l_{x}^{-2}$, and their height corresponds to a quantum of conductance, $2e^{2}/h$. \\ Considering the complex geometry of our  device (see Figure \ref{fig:fig2}(a)),  extracting a precise value for the conductance jumps from the experimental data is challenging, as we need to estimate and remove the contribution of the full channel (the long conducting path in series with the constriction) from the total conductance (Figure  \ref{fig:fig3}(b)). A quite good approximation is provided by the following method. 
First, we compute the resistance of the constriction, $R_{QPC}$, by subtracting from the total resistance the component of the full channel (the reservoirs), computed by  rescaling the total resistance to the reservoir length, $l_{channel}$, i.e.: 
\begin{equation}\label{eq:rescale}
 R_{QPC}=R_{tot}\cdot(1-l_{channel}/l_{tot}),
\end{equation}\\  
where  $R_{tot}$ is the total resistance of the device, and $l_{tot}$ the total length of the channel (full path and constriction region). We fixed $l_{channel}$ and $l_{tot}$ to 147.5~$\mu$m and  150~$\mu$m, respectively. The value of $l_{tot}$ is the nominal length of the pattern of our device, while $l_{channel}$ was chosen under the assumption that the insulating islands occupy approximately 2.5~$\mu m$ of the total channel length. Finally, from $R_{QPC}$, we computed the conductance of the constriction, obtaining a value that matches steps of magnitude $2e^{2}/h$.\\ 
In order to compare the experimental data  with the calcuations we need to estimate the scaling factor between the back-gate, $V_{g}$, and the electrostatic potential $V_{0}$. Indeed, these two quantities are not equivalent:  while  $V_{g}$ is the voltage  applied to the  back-electrode, $V_{0}$ is the electrostatic potential related to gate-induced Fermi energy shift. In order to convert the gate voltage into the electrostatic potential at the interface, we analyzed the effect of $\Delta V_{0}$ on a simple parabolic conduction band, with an electron effective mass of $2.2\cdot m$ \cite{fete_rashba_2012}. For a variation $\Delta V_{0}=0.12~\mathrm{meV}$ we computed a carrier density variation, $\Delta n_{\mathrm{2D}}$, of $1.1\cdot10^{11}~\mathrm{cm^{-2}}$. Since such  $\Delta n_{\mathrm{2D}}$ is obtained in our devices by $\Delta V_{g}$ of $\sim$1.4~V \cite{phdfete_2014} \footnote{We note that this value has been extracted from analysis performed on wider devices (later size above 100~$\mu$m), although  the field focusing occurring in narrow ones enables to achieve the same $\Delta n_{\mathrm{2D}}$ using a lower backgate voltage  \cite{rakhmilevitch_anomalous_2013}}, we estimate a scaling factor $\Delta V_{g}/\Delta V_{0}$ of 1.2$\cdot10^{4}$. 
As a result of this analysis,  we find a good agreement between the experimental and calculated curves (cf. Figure \ref{fig:fig6}). We note that the deviations of the experimental points from the calculated curves might have multiple origins. First of all, the real geometry of the constriction might deviate from the perfectly parabolic model in Equation \ref{eq:SP}. Secondly, there might be spurious effects stemming from other constrictions present in our device (see Fig. \ref{fig:fig2}).\\ 
If we extend our model to higher temperatures, we find that the calculated conductance steps are smeared out at 100~mK. Experimentally, the quantization seems to persist up to $1~$K, as visible in Figure \ref{fig:fig4}(a).  This could be due to a larger energy separation, $\hbar \omega_{y}$, between the conductance level due to a smaller channel size.  For example, if we consider a QPC with $l_{x}=8$~nm and $l_{y}=6$~nm, we find conductance steps similar to the experimental ones at 50~mK that remain essentially unchanged up to 1~K.
Therefore the real dimension of the constriction giving rise to this effect remains elusive.\\
While the QPC model ultimately gives reason for the insulating behavior observed in Figure \ref{fig:fig1}(f), it cannot reproduce the complete behavior of the resistance in temperature, since it completely neglects a plethora of phenomena occurring at higher temperature, such as electron-phonon and electron-electron interaction, that are expected to be the dominant contribution to the resistance in this temperature range.

\section{Discussion}
\begin{figure*}
 \begin{center}
  \includegraphics[width=1.2\columnwidth]{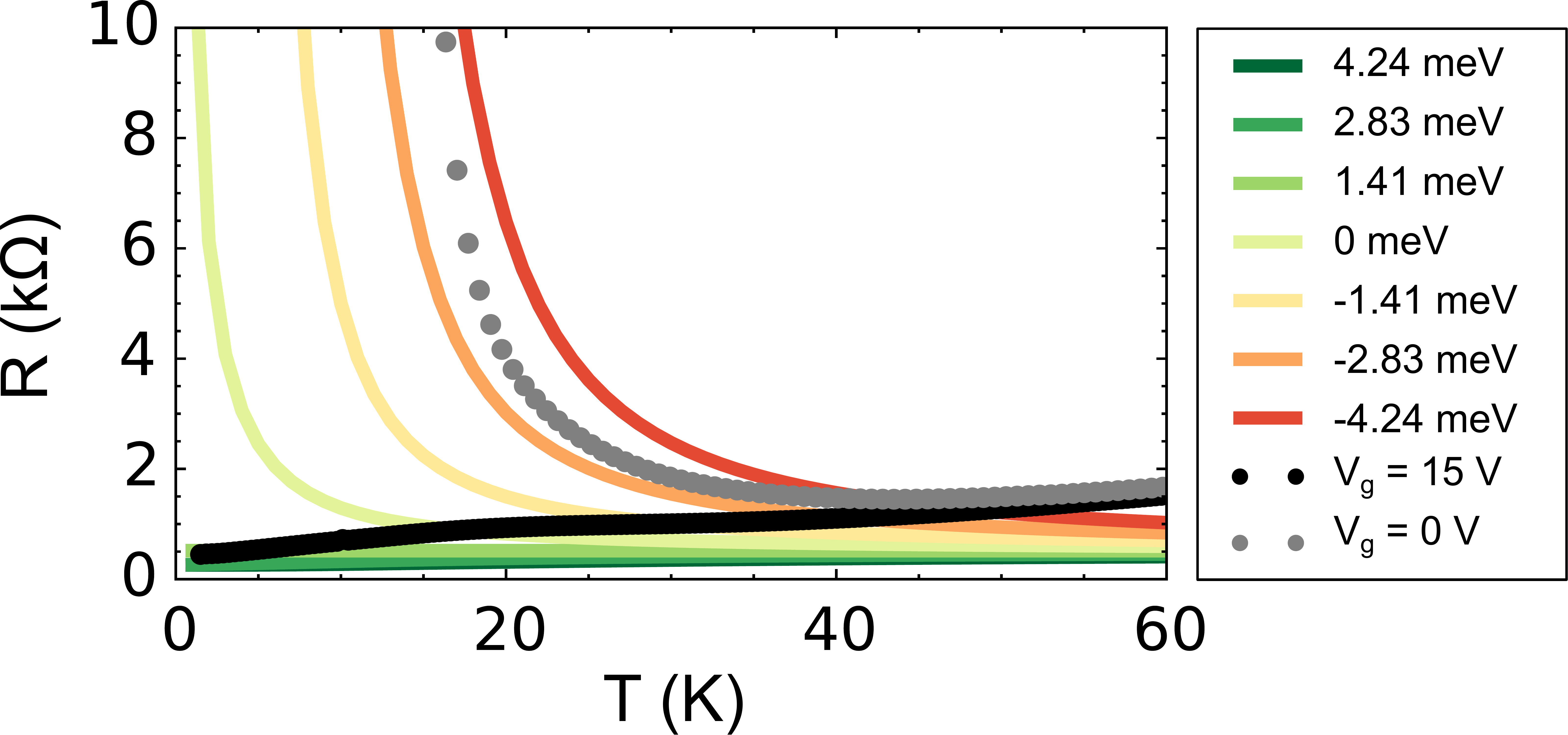} \\
 \end{center}
\caption{\label{fig:fig_disc} Calculated (solid lines) and experimental values (points) of the constriction resistance as a function of temperature. The calculations have been performed using the model and the parameters detailed in Section IV, with $l_{x}$=350~nm and $l_{y}$=35~nm. The values of the electrostatic potential $V_{0}$ used in the calculations are reported in the legend.}
\end{figure*}

The analysis presented in the previous section shows that the conductance jumps observed  in our device as a function of the back-gate result from the quantization of the transverse electronic states in a QPC. We also followed experimentally the evolution of these plateaus in temperature and magnetic field. 
While the temperature dependence of the conductance quantization corresponds to what is expected for a QPC \cite{mesoscphys, semicondnanos}, 
the behavior in magnetic field is less trivial to understand. The saddle-point model predicts a modification of the plateaus when a magnetic field is applied to the system, as a result of the interplay between the quantization of the energy levels induced by the confinement, $\omega_{x,y}$, and by the magnetic field ($\omega_{c}=|eB|/m$, where $e$ is the electron charge,  $m$ its mass,
and $B$ the magnetic field)  \cite{buttiker_quantized_1990}. The inclusion of orbital effects in our model would lead to sharper conductance steps as the magnetic field is increased 
\cite{buttiker_quantized_1990}.\\
In our experiments, by increasing the field to 8~T, we observe an opposite effect, since the magnetic field suppresses the conductance plateaus (see Figure \ref{fig:fig4}(b)).\\ A recent report by Jouan \textit{et al.} \cite{jouan_QPC_2020} on transport measurements in magnetic field, realized in a gate-controlled QPC at the LaAlO$_{3}$/SrTiO$_{3}$ interface, revealed the emergence of half-integer quantum conductance steps, 0.5($2e^{2}$/h), at 6~T. They attribute this effect to the Zeeman splitting of the spin-degenerated QPC levels. In our system we could not observe  the emergence of such additional steps. The behavior of our device might originate from additional effects of the magnetic field on the electronic reservoirs, characterized by the interplay between spin-orbit coupling, Zeeman splitting and lateral confinement provided by the lithographically defined channel, whose description goes well beyond the simplistic saddle-point model for the QPC.\\
The presence of quantum tunnelling barriers in conducting channels is a good candidate to explain the crossover between  metallic and  insulating behavior observed in our devices. In order to analyse this scenario, we used the model presented in Section IV to compute the temperature behavior of the QPC resistance for different values of the electrostatic potential (see Figure \ref{fig:fig_disc}). Despite the limit of this calculation, that neglects any effect occurring at high temperature such as electron-phonon and electron-electron coupling, we observe that,  for $V_{0}\leq 0$, the resistance of the system diverges below $\sim$30~K (Figure \ref{fig:fig_disc}). The metallic behavior is restored only at positive values of the electrostatic potential, similarly to what we observed experimentally (see for comparison the experimental points in Figure \ref{fig:fig_disc}). \\
As discussed in the experimental section,   Figure \ref{fig:fig1} shows that 
AFM-written wires become insulating at 
temperatures which are comparable to the ones observed in the device patterned with amorphous SrTiO$_{3}$ ($\sim$ 40~K). 
One may argue that this analogy is due to the presence of local constrictions in this electrically defined channels whose precise charge profile remains unknown. It is worth noting that, in this case, other phenomena could be equally responsible for the  insulating behavior at low temperatures: the presence of multiple constrictions could act as an effective disordered potential leading to Anderson localization \cite{anderson_absence_1958, abrahams_scaling_1979, lagendijk_fifty_2009}, a phenomenon which is robust to the presence of repulsive Coulomb interactions among electrons \cite{giamarchi_anderson_1988}. Alternatively, the formation of consecutive barriers could lead to the formation of well isolated charge puddles acting as quantum-dots responsible for Coulomb blockade effects \cite{aleiner_quantum_2002}. Further efforts in controlling the shape of thin and homogeneous wires would be important to clarify this issue. \\
In conclusion, this work, aiming at exploring the electric transport properties of  nanoscale devices realized at the LaAlO$_{3}$/SrTiO$_{3}$ interface,  shows that the lateral size of a conducting path has a striking effect on the resistance behavior as a function of temperature. Studying devices realized using the AFM-writing technique or with conventional photolithography, both with a lateral confinement of the 2DES lower than $\sim$100~nm, we witnessed a crossover from a metallic to an insulating behavior at  $\sim$50~K. The comparison between experiments and model calculations reveals that this behavior could be understood by considering the presence of tunnel barriers acting as QPCs along the conducting path. While these findings will be useful for the interpretation of past and future experiments performed on such oxide 2D nanoscale devices, it will be of high interest studying the confinement effects in future devices, where superconductivity is present.

\color{black}

\section{Acknowledgements}
The authors wish to thank Marco Lopes and Sébastien Muller for technical support. This work was supported by the Swiss National Science Foundation through Division II (project 200020-179155), M.F by the FNS/SNF Ambizione Grant PZ00P2\textunderscore174038. The research leading to these results has received funding from the European Research Council under the European Union’s Seventh Framework Program (FP7/2007-2013)/ERC Grant Agreement 319286 Q-MAC).

\bibliography{biblio}

\begin{thebibliography}{43}
\expandafter\ifx\csname natexlab\endcsname\relax\def\natexlab#1{#1}\fi
\expandafter\ifx\csname bibnamefont\endcsname\relax
  \def\bibnamefont#1{#1}\fi
\expandafter\ifx\csname bibfnamefont\endcsname\relax
  \def\bibfnamefont#1{#1}\fi
\expandafter\ifx\csname citenamefont\endcsname\relax
  \def\citenamefont#1{#1}\fi
\expandafter\ifx\csname url\endcsname\relax
  \def\url#1{\texttt{#1}}\fi
\expandafter\ifx\csname urlprefix\endcsname\relax\def\urlprefix{URL }\fi
\providecommand{\bibinfo}[2]{#2}
\providecommand{\eprint}[2][]{\url{#2}}

\bibitem[{\citenamefont{Ohtomo and Hwang}(2004)}]{ohtomo_2004}
\bibinfo{author}{\bibfnamefont{A.}~\bibnamefont{Ohtomo}} \bibnamefont{and}
  \bibinfo{author}{\bibfnamefont{H.~Y.} \bibnamefont{Hwang}},
  \bibinfo{journal}{Nature} \textbf{\bibinfo{volume}{427}},
  \bibinfo{pages}{423} (\bibinfo{year}{2004}), ISSN \bibinfo{issn}{0028-0836},
  \urlprefix\url{http://www.nature.com/nature/journal/v427/n6973/full/nature02308.html}.

\bibitem[{\citenamefont{Reyren et~al.}(2007)\citenamefont{Reyren, Thiel,
  Caviglia, Kourkoutis, Hammerl, Richter, Schneider, Kopp, R\"uetschi, Jaccard
  et~al.}}]{reyren_superconducting_2007}
\bibinfo{author}{\bibfnamefont{N.}~\bibnamefont{Reyren}},
  \bibinfo{author}{\bibfnamefont{S.}~\bibnamefont{Thiel}},
  \bibinfo{author}{\bibfnamefont{A.~D.} \bibnamefont{Caviglia}},
  \bibinfo{author}{\bibfnamefont{L.~F.} \bibnamefont{Kourkoutis}},
  \bibinfo{author}{\bibfnamefont{G.}~\bibnamefont{Hammerl}},
  \bibinfo{author}{\bibfnamefont{C.}~\bibnamefont{Richter}},
  \bibinfo{author}{\bibfnamefont{C.~W.} \bibnamefont{Schneider}},
  \bibinfo{author}{\bibfnamefont{T.}~\bibnamefont{Kopp}},
  \bibinfo{author}{\bibfnamefont{A.-S.} \bibnamefont{R\"uetschi}},
  \bibinfo{author}{\bibfnamefont{D.}~\bibnamefont{Jaccard}},
  \bibnamefont{et~al.}, \bibinfo{journal}{Science}
  \textbf{\bibinfo{volume}{317}}, \bibinfo{pages}{1196} (\bibinfo{year}{2007}),
  ISSN \bibinfo{issn}{0036-8075, 1095-9203},
  \urlprefix\url{http://www.sciencemag.org/content/317/5842/1196}.

\bibitem[{\citenamefont{Caviglia et~al.}(2008)\citenamefont{Caviglia, Gariglio,
  Reyren, Jaccard, Schneider, Gabay, Thiel, Hammerl, Mannhart, and
  Triscone}}]{caviglia_electric_2008}
\bibinfo{author}{\bibfnamefont{A.~D.} \bibnamefont{Caviglia}},
  \bibinfo{author}{\bibfnamefont{S.}~\bibnamefont{Gariglio}},
  \bibinfo{author}{\bibfnamefont{N.}~\bibnamefont{Reyren}},
  \bibinfo{author}{\bibfnamefont{D.}~\bibnamefont{Jaccard}},
  \bibinfo{author}{\bibfnamefont{T.}~\bibnamefont{Schneider}},
  \bibinfo{author}{\bibfnamefont{M.}~\bibnamefont{Gabay}},
  \bibinfo{author}{\bibfnamefont{S.}~\bibnamefont{Thiel}},
  \bibinfo{author}{\bibfnamefont{G.}~\bibnamefont{Hammerl}},
  \bibinfo{author}{\bibfnamefont{J.}~\bibnamefont{Mannhart}}, \bibnamefont{and}
  \bibinfo{author}{\bibfnamefont{J.-M.} \bibnamefont{Triscone}},
  \bibinfo{journal}{Nature} \textbf{\bibinfo{volume}{456}},
  \bibinfo{pages}{624} (\bibinfo{year}{2008}), ISSN \bibinfo{issn}{0028-0836},
  \urlprefix\url{http://www.nature.com/nature/journal/v456/n7222/abs/nature07576.html}.

\bibitem[{\citenamefont{Caviglia et~al.}(2010)\citenamefont{Caviglia, Gabay,
  Gariglio, Reyren, Cancellieri, and Triscone}}]{caviglia_tunable_2010}
\bibinfo{author}{\bibfnamefont{A.~D.} \bibnamefont{Caviglia}},
  \bibinfo{author}{\bibfnamefont{M.}~\bibnamefont{Gabay}},
  \bibinfo{author}{\bibfnamefont{S.}~\bibnamefont{Gariglio}},
  \bibinfo{author}{\bibfnamefont{N.}~\bibnamefont{Reyren}},
  \bibinfo{author}{\bibfnamefont{C.}~\bibnamefont{Cancellieri}},
  \bibnamefont{and} \bibinfo{author}{\bibfnamefont{J.-M.}
  \bibnamefont{Triscone}}, \bibinfo{journal}{Physical Review Letters}
  \textbf{\bibinfo{volume}{104}}, \bibinfo{pages}{126803}
  (\bibinfo{year}{2010}),
  \urlprefix\url{http://link.aps.org/doi/10.1103/PhysRevLett.104.126803}.

\bibitem[{\citenamefont{Lesne et~al.}(2016)\citenamefont{Lesne, Fu, Oyarzun,
  Rojas-Sánchez, Vaz, Naganuma, Sicoli, Attané, Jamet, Jacquet
  et~al.}}]{lesne_highly_2016}
\bibinfo{author}{\bibfnamefont{E.}~\bibnamefont{Lesne}},
  \bibinfo{author}{\bibfnamefont{Y.}~\bibnamefont{Fu}},
  \bibinfo{author}{\bibfnamefont{S.}~\bibnamefont{Oyarzun}},
  \bibinfo{author}{\bibfnamefont{J.~C.} \bibnamefont{Rojas-Sánchez}},
  \bibinfo{author}{\bibfnamefont{D.~C.} \bibnamefont{Vaz}},
  \bibinfo{author}{\bibfnamefont{H.}~\bibnamefont{Naganuma}},
  \bibinfo{author}{\bibfnamefont{G.}~\bibnamefont{Sicoli}},
  \bibinfo{author}{\bibfnamefont{J.-P.} \bibnamefont{Attané}},
  \bibinfo{author}{\bibfnamefont{M.}~\bibnamefont{Jamet}},
  \bibinfo{author}{\bibfnamefont{E.}~\bibnamefont{Jacquet}},
  \bibnamefont{et~al.}, \bibinfo{journal}{Nature Materials}
  (\bibinfo{year}{2016}), ISSN \bibinfo{issn}{1476-1122},
  \urlprefix\url{http://www.nature.com/nmat/journal/vaop/ncurrent/full/nmat4726.html}.

\bibitem[{\citenamefont{Chauleau et~al.}(2016)\citenamefont{Chauleau, Boselli,
  Gariglio, Weil, Loubens, Triscone, and Viret}}]{chauleau_2016}
\bibinfo{author}{\bibfnamefont{J.-Y.} \bibnamefont{Chauleau}},
  \bibinfo{author}{\bibfnamefont{M.}~\bibnamefont{Boselli}},
  \bibinfo{author}{\bibfnamefont{S.}~\bibnamefont{Gariglio}},
  \bibinfo{author}{\bibfnamefont{R.}~\bibnamefont{Weil}},
  \bibinfo{author}{\bibfnamefont{G.~d.} \bibnamefont{Loubens}},
  \bibinfo{author}{\bibfnamefont{J.-M.} \bibnamefont{Triscone}},
  \bibnamefont{and} \bibinfo{author}{\bibfnamefont{M.}~\bibnamefont{Viret}},
  \bibinfo{journal}{EPL} \textbf{\bibinfo{volume}{116}}, \bibinfo{pages}{17006}
  (\bibinfo{year}{2016}), ISSN \bibinfo{issn}{0295-5075, 1286-4854},
  \urlprefix\url{http://dx.doi.org/10.1209/0295-5075/116/17006}.

\bibitem[{\citenamefont{Song et~al.}(2017)\citenamefont{Song, Zhang, Su, Yuan,
  Chen, Xing, Shi, Sun, and Han}}]{song_observation_2017}
\bibinfo{author}{\bibfnamefont{Q.}~\bibnamefont{Song}},
  \bibinfo{author}{\bibfnamefont{H.}~\bibnamefont{Zhang}},
  \bibinfo{author}{\bibfnamefont{T.}~\bibnamefont{Su}},
  \bibinfo{author}{\bibfnamefont{W.}~\bibnamefont{Yuan}},
  \bibinfo{author}{\bibfnamefont{Y.}~\bibnamefont{Chen}},
  \bibinfo{author}{\bibfnamefont{W.}~\bibnamefont{Xing}},
  \bibinfo{author}{\bibfnamefont{J.}~\bibnamefont{Shi}},
  \bibinfo{author}{\bibfnamefont{J.}~\bibnamefont{Sun}}, \bibnamefont{and}
  \bibinfo{author}{\bibfnamefont{W.}~\bibnamefont{Han}},
  \bibinfo{journal}{Science Advances} \textbf{\bibinfo{volume}{3}},
  \bibinfo{pages}{e1602312} (\bibinfo{year}{2017}), ISSN
  \bibinfo{issn}{2375-2548},
  \urlprefix\url{http://advances.sciencemag.org/content/3/3/e1602312}.

\bibitem[{\citenamefont{Vaz et~al.}(2019)\citenamefont{Vaz, Noël, Johansson,
  Göbel, Bruno, Singh, McKeown-Walker, Trier, Vicente-Arche, Sander
  et~al.}}]{vaz_topology_2019}
\bibinfo{author}{\bibfnamefont{D.~C.} \bibnamefont{Vaz}},
  \bibinfo{author}{\bibfnamefont{P.}~\bibnamefont{Noël}},
  \bibinfo{author}{\bibfnamefont{A.}~\bibnamefont{Johansson}},
  \bibinfo{author}{\bibfnamefont{B.}~\bibnamefont{Göbel}},
  \bibinfo{author}{\bibfnamefont{F.~Y.} \bibnamefont{Bruno}},
  \bibinfo{author}{\bibfnamefont{G.}~\bibnamefont{Singh}},
  \bibinfo{author}{\bibfnamefont{S.}~\bibnamefont{McKeown-Walker}},
  \bibinfo{author}{\bibfnamefont{F.}~\bibnamefont{Trier}},
  \bibinfo{author}{\bibfnamefont{L.~M.} \bibnamefont{Vicente-Arche}},
  \bibinfo{author}{\bibfnamefont{A.}~\bibnamefont{Sander}},
  \bibnamefont{et~al.}, \bibinfo{journal}{Nature Materials}
  \textbf{\bibinfo{volume}{18}}, \bibinfo{pages}{1187} (\bibinfo{year}{2019}),
  ISSN \bibinfo{issn}{1476-4660},
  \urlprefix\url{https://www.nature.com/articles/s41563-019-0467-4}.

\bibitem[{\citenamefont{Trier et~al.}(2020)\citenamefont{Trier, Vaz, Bruneel,
  Noël, Fert, Vila, Attané, Barthélémy, Gabay, Jaffrès
  et~al.}}]{trier_electric-field_2020}
\bibinfo{author}{\bibfnamefont{F.}~\bibnamefont{Trier}},
  \bibinfo{author}{\bibfnamefont{D.~C.} \bibnamefont{Vaz}},
  \bibinfo{author}{\bibfnamefont{P.}~\bibnamefont{Bruneel}},
  \bibinfo{author}{\bibfnamefont{P.}~\bibnamefont{Noël}},
  \bibinfo{author}{\bibfnamefont{A.}~\bibnamefont{Fert}},
  \bibinfo{author}{\bibfnamefont{L.}~\bibnamefont{Vila}},
  \bibinfo{author}{\bibfnamefont{J.-P.} \bibnamefont{Attané}},
  \bibinfo{author}{\bibfnamefont{A.}~\bibnamefont{Barthélémy}},
  \bibinfo{author}{\bibfnamefont{M.}~\bibnamefont{Gabay}},
  \bibinfo{author}{\bibfnamefont{H.}~\bibnamefont{Jaffrès}},
  \bibnamefont{et~al.}, \bibinfo{journal}{Nano Letters}
  \textbf{\bibinfo{volume}{20}}, \bibinfo{pages}{395} (\bibinfo{year}{2020}),
  ISSN \bibinfo{issn}{1530-6984},
  \urlprefix\url{https://doi.org/10.1021/acs.nanolett.9b04079}.

\bibitem[{\citenamefont{Rakhmilevitch et~al.}(2010)\citenamefont{Rakhmilevitch,
  Ben~Shalom, Eshkol, Tsukernik, Palevski, and
  Dagan}}]{rakhmilevitch_phase_2010}
\bibinfo{author}{\bibfnamefont{D.}~\bibnamefont{Rakhmilevitch}},
  \bibinfo{author}{\bibfnamefont{M.}~\bibnamefont{Ben~Shalom}},
  \bibinfo{author}{\bibfnamefont{M.}~\bibnamefont{Eshkol}},
  \bibinfo{author}{\bibfnamefont{A.}~\bibnamefont{Tsukernik}},
  \bibinfo{author}{\bibfnamefont{A.}~\bibnamefont{Palevski}}, \bibnamefont{and}
  \bibinfo{author}{\bibfnamefont{Y.}~\bibnamefont{Dagan}},
  \bibinfo{journal}{Physical Review B} \textbf{\bibinfo{volume}{82}},
  \bibinfo{pages}{235119} (\bibinfo{year}{2010}),
  \urlprefix\url{http://link.aps.org/doi/10.1103/PhysRevB.82.235119}.

\bibitem[{\citenamefont{Stornaiuolo et~al.}(2012)\citenamefont{Stornaiuolo,
  Gariglio, Couto, Fête, Caviglia, Seyfarth, Jaccard, Morpurgo, and
  Triscone}}]{stornaiuolo_plane_2012}
\bibinfo{author}{\bibfnamefont{D.}~\bibnamefont{Stornaiuolo}},
  \bibinfo{author}{\bibfnamefont{S.}~\bibnamefont{Gariglio}},
  \bibinfo{author}{\bibfnamefont{N.~J.~G.} \bibnamefont{Couto}},
  \bibinfo{author}{\bibfnamefont{A.}~\bibnamefont{Fête}},
  \bibinfo{author}{\bibfnamefont{A.~D.} \bibnamefont{Caviglia}},
  \bibinfo{author}{\bibfnamefont{G.}~\bibnamefont{Seyfarth}},
  \bibinfo{author}{\bibfnamefont{D.}~\bibnamefont{Jaccard}},
  \bibinfo{author}{\bibfnamefont{A.~F.} \bibnamefont{Morpurgo}},
  \bibnamefont{and} \bibinfo{author}{\bibfnamefont{J.-M.}
  \bibnamefont{Triscone}}, \bibinfo{journal}{Applied Physics Letters}
  \textbf{\bibinfo{volume}{101}}, \bibinfo{pages}{222601}
  (\bibinfo{year}{2012}), ISSN \bibinfo{issn}{0003-6951, 1077-3118},
  \urlprefix\url{http://scitation.aip.org/content/aip/journal/apl/101/22/10.1063/1.4768936}.

\bibitem[{\citenamefont{Aurino et~al.}(2013)\citenamefont{Aurino, Kalabukhov,
  Tuzla, Olsson, Claeson, and Winkler}}]{aurino_nano-patterning_2013}
\bibinfo{author}{\bibfnamefont{P.~P.} \bibnamefont{Aurino}},
  \bibinfo{author}{\bibfnamefont{A.}~\bibnamefont{Kalabukhov}},
  \bibinfo{author}{\bibfnamefont{N.}~\bibnamefont{Tuzla}},
  \bibinfo{author}{\bibfnamefont{E.}~\bibnamefont{Olsson}},
  \bibinfo{author}{\bibfnamefont{T.}~\bibnamefont{Claeson}}, \bibnamefont{and}
  \bibinfo{author}{\bibfnamefont{D.}~\bibnamefont{Winkler}},
  \bibinfo{journal}{Applied Physics Letters} \textbf{\bibinfo{volume}{102}},
  \bibinfo{pages}{201610} (\bibinfo{year}{2013}), ISSN
  \bibinfo{issn}{0003-6951, 1077-3118},
  \urlprefix\url{http://scitation.aip.org/content/aip/journal/apl/102/20/10.1063/1.4807785}.

\bibitem[{\citenamefont{Goswami et~al.}(2016)\citenamefont{Goswami,
  Mulazimoglu, Monteiro, Wölbing, Koelle, Kleiner, Blanter, Vandersypen, and
  Caviglia}}]{goswami_quantum_2016}
\bibinfo{author}{\bibfnamefont{S.}~\bibnamefont{Goswami}},
  \bibinfo{author}{\bibfnamefont{E.}~\bibnamefont{Mulazimoglu}},
  \bibinfo{author}{\bibfnamefont{A.~M. R. V.~L.} \bibnamefont{Monteiro}},
  \bibinfo{author}{\bibfnamefont{R.}~\bibnamefont{Wölbing}},
  \bibinfo{author}{\bibfnamefont{D.}~\bibnamefont{Koelle}},
  \bibinfo{author}{\bibfnamefont{R.}~\bibnamefont{Kleiner}},
  \bibinfo{author}{\bibfnamefont{Y.~M.} \bibnamefont{Blanter}},
  \bibinfo{author}{\bibfnamefont{L.~M.~K.} \bibnamefont{Vandersypen}},
  \bibnamefont{and} \bibinfo{author}{\bibfnamefont{A.~D.}
  \bibnamefont{Caviglia}}, \bibinfo{journal}{Nature Nanotechnology}
  \textbf{\bibinfo{volume}{11}}, \bibinfo{pages}{861} (\bibinfo{year}{2016}),
  ISSN \bibinfo{issn}{1748-3387, 1748-3395},
  \urlprefix\url{http://www.nature.com/articles/nnano.2016.112}.

\bibitem[{\citenamefont{Cen et~al.}(2008)\citenamefont{Cen, Thiel, Hammerl,
  Schneider, Andersen, Hellberg, Mannhart, and Levy}}]{cen_nanoscale_2008}
\bibinfo{author}{\bibfnamefont{C.}~\bibnamefont{Cen}},
  \bibinfo{author}{\bibfnamefont{S.}~\bibnamefont{Thiel}},
  \bibinfo{author}{\bibfnamefont{G.}~\bibnamefont{Hammerl}},
  \bibinfo{author}{\bibfnamefont{C.~W.} \bibnamefont{Schneider}},
  \bibinfo{author}{\bibfnamefont{K.~E.} \bibnamefont{Andersen}},
  \bibinfo{author}{\bibfnamefont{C.~S.} \bibnamefont{Hellberg}},
  \bibinfo{author}{\bibfnamefont{J.}~\bibnamefont{Mannhart}}, \bibnamefont{and}
  \bibinfo{author}{\bibfnamefont{J.}~\bibnamefont{Levy}},
  \bibinfo{journal}{Nature Materials} \textbf{\bibinfo{volume}{7}},
  \bibinfo{pages}{298} (\bibinfo{year}{2008}), ISSN \bibinfo{issn}{1476-1122},
  \urlprefix\url{http://www.nature.com/nmat/journal/v7/n4/abs/nmat2136.html}.

\bibitem[{\citenamefont{Cen et~al.}(2009)\citenamefont{Cen, Thiel, Mannhart,
  and Levy}}]{cen_oxide_2009}
\bibinfo{author}{\bibfnamefont{C.}~\bibnamefont{Cen}},
  \bibinfo{author}{\bibfnamefont{S.}~\bibnamefont{Thiel}},
  \bibinfo{author}{\bibfnamefont{J.}~\bibnamefont{Mannhart}}, \bibnamefont{and}
  \bibinfo{author}{\bibfnamefont{J.}~\bibnamefont{Levy}},
  \bibinfo{journal}{Science} \textbf{\bibinfo{volume}{323}},
  \bibinfo{pages}{1026} (\bibinfo{year}{2009}), ISSN \bibinfo{issn}{0036-8075,
  1095-9203}, \urlprefix\url{http://www.sciencemag.org/content/323/5917/1026}.

\bibitem[{\citenamefont{Goswami et~al.}(2015)\citenamefont{Goswami,
  Mulazimoglu, Vandersypen, and Caviglia}}]{goswami_nanoscale_2015}
\bibinfo{author}{\bibfnamefont{S.}~\bibnamefont{Goswami}},
  \bibinfo{author}{\bibfnamefont{E.}~\bibnamefont{Mulazimoglu}},
  \bibinfo{author}{\bibfnamefont{L.~M.~K.} \bibnamefont{Vandersypen}},
  \bibnamefont{and} \bibinfo{author}{\bibfnamefont{A.~D.}
  \bibnamefont{Caviglia}}, \bibinfo{journal}{Nano Letters}
  \textbf{\bibinfo{volume}{15}}, \bibinfo{pages}{2627} (\bibinfo{year}{2015}),
  ISSN \bibinfo{issn}{1530-6984},
  \urlprefix\url{http://dx.doi.org/10.1021/acs.nanolett.5b00216}.

\bibitem[{\citenamefont{Monteiro et~al.}(2017)\citenamefont{Monteiro,
  Groenendijk, Manca, Mulazimoglu, Goswami, Blanter, Vandersypen, and
  Caviglia}}]{monteiro_side_2017}
\bibinfo{author}{\bibfnamefont{A.~M. R. V.~L.} \bibnamefont{Monteiro}},
  \bibinfo{author}{\bibfnamefont{D.~J.} \bibnamefont{Groenendijk}},
  \bibinfo{author}{\bibfnamefont{N.}~\bibnamefont{Manca}},
  \bibinfo{author}{\bibfnamefont{E.}~\bibnamefont{Mulazimoglu}},
  \bibinfo{author}{\bibfnamefont{S.}~\bibnamefont{Goswami}},
  \bibinfo{author}{\bibfnamefont{Y.}~\bibnamefont{Blanter}},
  \bibinfo{author}{\bibfnamefont{L.~M.~K.} \bibnamefont{Vandersypen}},
  \bibnamefont{and} \bibinfo{author}{\bibfnamefont{A.~D.}
  \bibnamefont{Caviglia}}, \bibinfo{journal}{Nano Letters}
  \textbf{\bibinfo{volume}{17}}, \bibinfo{pages}{715} (\bibinfo{year}{2017}),
  ISSN \bibinfo{issn}{1530-6984},
  \urlprefix\url{http://dx.doi.org/10.1021/acs.nanolett.6b03820}.

\bibitem[{\citenamefont{Stornaiuolo et~al.}(2017)\citenamefont{Stornaiuolo,
  Massarotti, Di~Capua, Lucignano, Pepe, Salluzzo, and
  Tafuri}}]{stornaiuolo_2017}
\bibinfo{author}{\bibfnamefont{D.}~\bibnamefont{Stornaiuolo}},
  \bibinfo{author}{\bibfnamefont{D.}~\bibnamefont{Massarotti}},
  \bibinfo{author}{\bibfnamefont{R.}~\bibnamefont{Di~Capua}},
  \bibinfo{author}{\bibfnamefont{P.}~\bibnamefont{Lucignano}},
  \bibinfo{author}{\bibfnamefont{G.~P.} \bibnamefont{Pepe}},
  \bibinfo{author}{\bibfnamefont{M.}~\bibnamefont{Salluzzo}}, \bibnamefont{and}
  \bibinfo{author}{\bibfnamefont{F.}~\bibnamefont{Tafuri}},
  \bibinfo{journal}{Physical Review B} \textbf{\bibinfo{volume}{95}},
  \bibinfo{pages}{140502} (\bibinfo{year}{2017}), ISSN
  \bibinfo{issn}{2469-9950, 2469-9969},
  \urlprefix\url{http://link.aps.org/doi/10.1103/PhysRevB.95.140502}.

\bibitem[{\citenamefont{Veazey et~al.}(2012)\citenamefont{Veazey, Cheng, Irvin,
  Cen, Bogorin, Bi, Huang, Bark, Ryu, Cho
  et~al.}}]{veazey_superconducting_2012}
\bibinfo{author}{\bibfnamefont{J.~P.} \bibnamefont{Veazey}},
  \bibinfo{author}{\bibfnamefont{G.}~\bibnamefont{Cheng}},
  \bibinfo{author}{\bibfnamefont{P.}~\bibnamefont{Irvin}},
  \bibinfo{author}{\bibfnamefont{C.}~\bibnamefont{Cen}},
  \bibinfo{author}{\bibfnamefont{D.~F.} \bibnamefont{Bogorin}},
  \bibinfo{author}{\bibfnamefont{F.}~\bibnamefont{Bi}},
  \bibinfo{author}{\bibfnamefont{M.}~\bibnamefont{Huang}},
  \bibinfo{author}{\bibfnamefont{C.-W.} \bibnamefont{Bark}},
  \bibinfo{author}{\bibfnamefont{S.}~\bibnamefont{Ryu}},
  \bibinfo{author}{\bibfnamefont{K.-H.} \bibnamefont{Cho}},
  \bibnamefont{et~al.}, \bibinfo{journal}{arXiv:1210.3606 [cond-mat]}
  (\bibinfo{year}{2012}), \urlprefix\url{http://arxiv.org/abs/1210.3606}.

\bibitem[{\citenamefont{Veazey et~al.}(2013)\citenamefont{Veazey, Cheng, Irvin,
  Cen, Bogorin, Bi, {Mengchen Huang}, Bark, Ryu, Cho
  et~al.}}]{veazey_oxide-based_2013}
\bibinfo{author}{\bibfnamefont{J.~P.} \bibnamefont{Veazey}},
  \bibinfo{author}{\bibfnamefont{G.}~\bibnamefont{Cheng}},
  \bibinfo{author}{\bibfnamefont{P.}~\bibnamefont{Irvin}},
  \bibinfo{author}{\bibfnamefont{C.}~\bibnamefont{Cen}},
  \bibinfo{author}{\bibfnamefont{D.~F.} \bibnamefont{Bogorin}},
  \bibinfo{author}{\bibfnamefont{F.}~\bibnamefont{Bi}},
  \bibinfo{author}{\bibnamefont{{Mengchen Huang}}},
  \bibinfo{author}{\bibfnamefont{C.-W.} \bibnamefont{Bark}},
  \bibinfo{author}{\bibfnamefont{S.}~\bibnamefont{Ryu}},
  \bibinfo{author}{\bibfnamefont{K.-H.} \bibnamefont{Cho}},
  \bibnamefont{et~al.}, \bibinfo{journal}{Nanotechnology}
  \textbf{\bibinfo{volume}{24}}, \bibinfo{pages}{375201}
  (\bibinfo{year}{2013}), ISSN \bibinfo{issn}{0957-4484},
  \urlprefix\url{http://stacks.iop.org/0957-4484/24/i=37/a=375201}.

\bibitem[{\citenamefont{Ron and Dagan}(2014)}]{ron_2014}
\bibinfo{author}{\bibfnamefont{A.}~\bibnamefont{Ron}} \bibnamefont{and}
  \bibinfo{author}{\bibfnamefont{Y.}~\bibnamefont{Dagan}},
  \bibinfo{journal}{Physical Review Letters} \textbf{\bibinfo{volume}{112}},
  \bibinfo{pages}{136801} (\bibinfo{year}{2014}),
  \urlprefix\url{http://link.aps.org/doi/10.1103/PhysRevLett.112.136801}.

\bibitem[{\citenamefont{Tomczyk et~al.}(2016)\citenamefont{Tomczyk, Cheng, Lee,
  Lu, Annadi, Veazey, Huang, Irvin, Ryu, Eom et~al.}}]{tomczyk_2016}
\bibinfo{author}{\bibfnamefont{M.}~\bibnamefont{Tomczyk}},
  \bibinfo{author}{\bibfnamefont{G.}~\bibnamefont{Cheng}},
  \bibinfo{author}{\bibfnamefont{H.}~\bibnamefont{Lee}},
  \bibinfo{author}{\bibfnamefont{S.}~\bibnamefont{Lu}},
  \bibinfo{author}{\bibfnamefont{A.}~\bibnamefont{Annadi}},
  \bibinfo{author}{\bibfnamefont{J.~P.} \bibnamefont{Veazey}},
  \bibinfo{author}{\bibfnamefont{M.}~\bibnamefont{Huang}},
  \bibinfo{author}{\bibfnamefont{P.}~\bibnamefont{Irvin}},
  \bibinfo{author}{\bibfnamefont{S.}~\bibnamefont{Ryu}},
  \bibinfo{author}{\bibfnamefont{C.-B.} \bibnamefont{Eom}},
  \bibnamefont{et~al.}, \bibinfo{journal}{Physical Review Letters}
  \textbf{\bibinfo{volume}{117}}, \bibinfo{pages}{096801}
  (\bibinfo{year}{2016}),
  \urlprefix\url{http://link.aps.org/doi/10.1103/PhysRevLett.117.096801}.

\bibitem[{\citenamefont{Thierschmann et~al.}(2018)\citenamefont{Thierschmann,
  Mulazimoglu, Manca, Goswami, Klapwijk, and Caviglia}}]{thierschmann_2018}
\bibinfo{author}{\bibfnamefont{H.}~\bibnamefont{Thierschmann}},
  \bibinfo{author}{\bibfnamefont{E.}~\bibnamefont{Mulazimoglu}},
  \bibinfo{author}{\bibfnamefont{N.}~\bibnamefont{Manca}},
  \bibinfo{author}{\bibfnamefont{S.}~\bibnamefont{Goswami}},
  \bibinfo{author}{\bibfnamefont{T.~M.} \bibnamefont{Klapwijk}},
  \bibnamefont{and} \bibinfo{author}{\bibfnamefont{A.~D.}
  \bibnamefont{Caviglia}}, \bibinfo{journal}{Nature Communications}
  \textbf{\bibinfo{volume}{9}}, \bibinfo{pages}{2276} (\bibinfo{year}{2018}),
  ISSN \bibinfo{issn}{2041-1723},
  \urlprefix\url{http://www.nature.com/articles/s41467-018-04657-z}.

\bibitem[{\citenamefont{Jouan et~al.}(2020)\citenamefont{Jouan, Singh, Lesne,
  Vaz, Bibes, Barthélémy, Ulysse, Stornaiuolo, Salluzzo, Hurand
  et~al.}}]{jouan_QPC_2020}
\bibinfo{author}{\bibfnamefont{A.}~\bibnamefont{Jouan}},
  \bibinfo{author}{\bibfnamefont{G.}~\bibnamefont{Singh}},
  \bibinfo{author}{\bibfnamefont{E.}~\bibnamefont{Lesne}},
  \bibinfo{author}{\bibfnamefont{D.~C.} \bibnamefont{Vaz}},
  \bibinfo{author}{\bibfnamefont{M.}~\bibnamefont{Bibes}},
  \bibinfo{author}{\bibfnamefont{A.}~\bibnamefont{Barthélémy}},
  \bibinfo{author}{\bibfnamefont{C.}~\bibnamefont{Ulysse}},
  \bibinfo{author}{\bibfnamefont{D.}~\bibnamefont{Stornaiuolo}},
  \bibinfo{author}{\bibfnamefont{M.}~\bibnamefont{Salluzzo}},
  \bibinfo{author}{\bibfnamefont{S.}~\bibnamefont{Hurand}},
  \bibnamefont{et~al.}, \bibinfo{journal}{Nature Electronics} pp.
  \bibinfo{pages}{201--206} (\bibinfo{year}{2020}), ISSN
  \bibinfo{issn}{2520-1131},
  \urlprefix\url{https://www.nature.com/articles/s41928-020-0383-2}.

\bibitem[{\citenamefont{Cheng et~al.}(2015)\citenamefont{Cheng, Tomczyk, Lu,
  Veazey, Huang, Irvin, Ryu, Lee, Eom, Hellberg et~al.}}]{cheng_electron_2015}
\bibinfo{author}{\bibfnamefont{G.}~\bibnamefont{Cheng}},
  \bibinfo{author}{\bibfnamefont{M.}~\bibnamefont{Tomczyk}},
  \bibinfo{author}{\bibfnamefont{S.}~\bibnamefont{Lu}},
  \bibinfo{author}{\bibfnamefont{J.~P.} \bibnamefont{Veazey}},
  \bibinfo{author}{\bibfnamefont{M.}~\bibnamefont{Huang}},
  \bibinfo{author}{\bibfnamefont{P.}~\bibnamefont{Irvin}},
  \bibinfo{author}{\bibfnamefont{S.}~\bibnamefont{Ryu}},
  \bibinfo{author}{\bibfnamefont{H.}~\bibnamefont{Lee}},
  \bibinfo{author}{\bibfnamefont{C.-B.} \bibnamefont{Eom}},
  \bibinfo{author}{\bibfnamefont{C.~S.} \bibnamefont{Hellberg}},
  \bibnamefont{et~al.}, \bibinfo{journal}{Nature}
  \textbf{\bibinfo{volume}{521}}, \bibinfo{pages}{196} (\bibinfo{year}{2015}),
  ISSN \bibinfo{issn}{0028-0836},
  \urlprefix\url{http://www.nature.com/nature/journal/v521/n7551/abs/nature14398.html}.

\bibitem[{\citenamefont{Aurino et~al.}(2016)\citenamefont{Aurino, Kalabukhov,
  Borgani, Haviland, Bauch, Lombardi, Claeson, and
  Winkler}}]{aurino_retention_2016}
\bibinfo{author}{\bibfnamefont{P.}~\bibnamefont{Aurino}},
  \bibinfo{author}{\bibfnamefont{A.}~\bibnamefont{Kalabukhov}},
  \bibinfo{author}{\bibfnamefont{R.}~\bibnamefont{Borgani}},
  \bibinfo{author}{\bibfnamefont{D.}~\bibnamefont{Haviland}},
  \bibinfo{author}{\bibfnamefont{T.}~\bibnamefont{Bauch}},
  \bibinfo{author}{\bibfnamefont{F.}~\bibnamefont{Lombardi}},
  \bibinfo{author}{\bibfnamefont{T.}~\bibnamefont{Claeson}}, \bibnamefont{and}
  \bibinfo{author}{\bibfnamefont{D.}~\bibnamefont{Winkler}},
  \bibinfo{journal}{Physical Review Applied} \textbf{\bibinfo{volume}{6}},
  \bibinfo{pages}{024011} (\bibinfo{year}{2016}), ISSN
  \bibinfo{issn}{2331-7019},
  \urlprefix\url{https://link.aps.org/doi/10.1103/PhysRevApplied.6.024011}.

\bibitem[{\citenamefont{Boselli et~al.}(2016)\citenamefont{Boselli, Li, Liu,
  Fête, Gariglio, and Triscone}}]{boselli_characterization_2016}
\bibinfo{author}{\bibfnamefont{M.}~\bibnamefont{Boselli}},
  \bibinfo{author}{\bibfnamefont{D.}~\bibnamefont{Li}},
  \bibinfo{author}{\bibfnamefont{W.}~\bibnamefont{Liu}},
  \bibinfo{author}{\bibfnamefont{A.}~\bibnamefont{Fête}},
  \bibinfo{author}{\bibfnamefont{S.}~\bibnamefont{Gariglio}}, \bibnamefont{and}
  \bibinfo{author}{\bibfnamefont{J.-M.} \bibnamefont{Triscone}},
  \bibinfo{journal}{Applied Physics Letters} \textbf{\bibinfo{volume}{108}},
  \bibinfo{pages}{061604} (\bibinfo{year}{2016}), ISSN
  \bibinfo{issn}{0003-6951, 1077-3118},
  \urlprefix\url{http://scitation.aip.org/content/aip/journal/apl/108/6/10.1063/1.4941817}.

\bibitem[{\citenamefont{Stornaiuolo et~al.}(2014)\citenamefont{Stornaiuolo,
  Gariglio, Fête, Gabay, Li, Massarotti, and
  Triscone}}]{stornaiuolo_weak_2014}
\bibinfo{author}{\bibfnamefont{D.}~\bibnamefont{Stornaiuolo}},
  \bibinfo{author}{\bibfnamefont{S.}~\bibnamefont{Gariglio}},
  \bibinfo{author}{\bibfnamefont{A.}~\bibnamefont{Fête}},
  \bibinfo{author}{\bibfnamefont{M.}~\bibnamefont{Gabay}},
  \bibinfo{author}{\bibfnamefont{D.}~\bibnamefont{Li}},
  \bibinfo{author}{\bibfnamefont{D.}~\bibnamefont{Massarotti}},
  \bibnamefont{and} \bibinfo{author}{\bibfnamefont{J.-M.}
  \bibnamefont{Triscone}}, \bibinfo{journal}{Physical Review B}
  \textbf{\bibinfo{volume}{90}}, \bibinfo{pages}{235426}
  (\bibinfo{year}{2014}),
  \urlprefix\url{http://link.aps.org/doi/10.1103/PhysRevB.90.235426}.

\bibitem[{\citenamefont{Li et~al.}(2018)\citenamefont{Li, Lemal, Gariglio, Wu,
  Fête, Boselli, Ghosez, and Triscone}}]{li_probing_2018}
\bibinfo{author}{\bibfnamefont{D.}~\bibnamefont{Li}},
  \bibinfo{author}{\bibfnamefont{S.}~\bibnamefont{Lemal}},
  \bibinfo{author}{\bibfnamefont{S.}~\bibnamefont{Gariglio}},
  \bibinfo{author}{\bibfnamefont{Z.}~\bibnamefont{Wu}},
  \bibinfo{author}{\bibfnamefont{A.}~\bibnamefont{Fête}},
  \bibinfo{author}{\bibfnamefont{M.}~\bibnamefont{Boselli}},
  \bibinfo{author}{\bibfnamefont{P.}~\bibnamefont{Ghosez}}, \bibnamefont{and}
  \bibinfo{author}{\bibfnamefont{J.-M.} \bibnamefont{Triscone}},
  \bibinfo{journal}{Advanced Science} \textbf{\bibinfo{volume}{5}},
  \bibinfo{pages}{1800242} (\bibinfo{year}{2018}), ISSN
  \bibinfo{issn}{2198-3844},
  \urlprefix\url{https://onlinelibrary.wiley.com/doi/abs/10.1002/advs.201800242}.

\bibitem[{\citenamefont{Luo et~al.}(2019)\citenamefont{Luo, Boselli, Poumirol,
  Ardizzone, Teyssier, van~der Marel, Gariglio, Triscone, and
  Kuzmenko}}]{luo_high_2019}
\bibinfo{author}{\bibfnamefont{W.}~\bibnamefont{Luo}},
  \bibinfo{author}{\bibfnamefont{M.}~\bibnamefont{Boselli}},
  \bibinfo{author}{\bibfnamefont{J.-M.} \bibnamefont{Poumirol}},
  \bibinfo{author}{\bibfnamefont{I.}~\bibnamefont{Ardizzone}},
  \bibinfo{author}{\bibfnamefont{J.}~\bibnamefont{Teyssier}},
  \bibinfo{author}{\bibfnamefont{D.}~\bibnamefont{van~der Marel}},
  \bibinfo{author}{\bibfnamefont{S.}~\bibnamefont{Gariglio}},
  \bibinfo{author}{\bibfnamefont{J.-M.} \bibnamefont{Triscone}},
  \bibnamefont{and} \bibinfo{author}{\bibfnamefont{A.~B.}
  \bibnamefont{Kuzmenko}}, \bibinfo{journal}{Nature Communications}
  \textbf{\bibinfo{volume}{10}}, \bibinfo{pages}{2774} (\bibinfo{year}{2019}),
  ISSN \bibinfo{issn}{2041-1723},
  \urlprefix\url{http://www.nature.com/articles/s41467-019-10672-5}.

\bibitem[{\citenamefont{Rakhmilevitch et~al.}(2013)\citenamefont{Rakhmilevitch,
  Neder, Shalom, Tsukernik, Karpovski, Dagan, and
  Palevski}}]{rakhmilevitch_anomalous_2013}
\bibinfo{author}{\bibfnamefont{D.}~\bibnamefont{Rakhmilevitch}},
  \bibinfo{author}{\bibfnamefont{I.}~\bibnamefont{Neder}},
  \bibinfo{author}{\bibfnamefont{M.~B.} \bibnamefont{Shalom}},
  \bibinfo{author}{\bibfnamefont{A.}~\bibnamefont{Tsukernik}},
  \bibinfo{author}{\bibfnamefont{M.}~\bibnamefont{Karpovski}},
  \bibinfo{author}{\bibfnamefont{Y.}~\bibnamefont{Dagan}}, \bibnamefont{and}
  \bibinfo{author}{\bibfnamefont{A.}~\bibnamefont{Palevski}},
  \bibinfo{journal}{Physical Review B} \textbf{\bibinfo{volume}{87}},
  \bibinfo{pages}{125409} (\bibinfo{year}{2013}), ISSN
  \bibinfo{issn}{1098-0121, 1550-235X},
  \urlprefix\url{https://link.aps.org/doi/10.1103/PhysRevB.87.125409}.

\bibitem[{\citenamefont{van Wees et~al.}(1988)\citenamefont{van Wees, van
  Houten, Beenakker, Williamson, Kouwenhoven, van~der Marel, and
  Foxon}}]{vanwees_1988}
\bibinfo{author}{\bibfnamefont{B.~J.} \bibnamefont{van Wees}},
  \bibinfo{author}{\bibfnamefont{H.}~\bibnamefont{van Houten}},
  \bibinfo{author}{\bibfnamefont{C.~W.~J.} \bibnamefont{Beenakker}},
  \bibinfo{author}{\bibfnamefont{J.~G.} \bibnamefont{Williamson}},
  \bibinfo{author}{\bibfnamefont{L.~P.} \bibnamefont{Kouwenhoven}},
  \bibinfo{author}{\bibfnamefont{D.}~\bibnamefont{van~der Marel}},
  \bibnamefont{and} \bibinfo{author}{\bibfnamefont{C.~T.} \bibnamefont{Foxon}},
  \bibinfo{journal}{Physical Review Letters} \textbf{\bibinfo{volume}{60}},
  \bibinfo{pages}{848} (\bibinfo{year}{1988}), ISSN \bibinfo{issn}{0031-9007},
  \urlprefix\url{https://link.aps.org/doi/10.1103/PhysRevLett.60.848}.

\bibitem[{\citenamefont{Akkermans and Montambaux}(2007)}]{mesoscphys}
\bibinfo{author}{\bibfnamefont{E.}~\bibnamefont{Akkermans}} \bibnamefont{and}
  \bibinfo{author}{\bibfnamefont{G.}~\bibnamefont{Montambaux}},
  \emph{\bibinfo{title}{Mesoscopic Physics of Electrons and Photons}}
  (\bibinfo{publisher}{Cambridge University Press}, \bibinfo{year}{2007}).

\bibitem[{\citenamefont{Ihn}(2010)}]{semicondnanos}
\bibinfo{author}{\bibfnamefont{T.}~\bibnamefont{Ihn}},
  \emph{\bibinfo{title}{Semiconductor Nanostructures. Quantum States and
  Electronic Transport}} (\bibinfo{publisher}{Oxford University Press},
  \bibinfo{year}{2010}).

\bibitem[{\citenamefont{B\"uttiker}(1990)}]{buttiker_quantized_1990}
\bibinfo{author}{\bibfnamefont{M.}~\bibnamefont{B\"uttiker}},
  \bibinfo{journal}{Physical Review B} \textbf{\bibinfo{volume}{41}},
  \bibinfo{pages}{7906} (\bibinfo{year}{1990}), ISSN \bibinfo{issn}{0163-1829,
  1095-3795},
  \urlprefix\url{https://link.aps.org/doi/10.1103/PhysRevB.41.7906}.

\bibitem[{\citenamefont{Fisher and Lee}(1981)}]{fisher_relation_1981}
\bibinfo{author}{\bibfnamefont{D.~S.} \bibnamefont{Fisher}} \bibnamefont{and}
  \bibinfo{author}{\bibfnamefont{P.~A.} \bibnamefont{Lee}},
  \bibinfo{journal}{Physical Review B} \textbf{\bibinfo{volume}{23}},
  \bibinfo{pages}{6851} (\bibinfo{year}{1981}), ISSN \bibinfo{issn}{0163-1829},
  \urlprefix\url{https://link.aps.org/doi/10.1103/PhysRevB.23.6851}.

\bibitem[{\citenamefont{Fête et~al.}(2012)\citenamefont{Fête, Gariglio,
  Caviglia, Triscone, and Gabay}}]{fete_rashba_2012}
\bibinfo{author}{\bibfnamefont{A.}~\bibnamefont{Fête}},
  \bibinfo{author}{\bibfnamefont{S.}~\bibnamefont{Gariglio}},
  \bibinfo{author}{\bibfnamefont{A.~D.} \bibnamefont{Caviglia}},
  \bibinfo{author}{\bibfnamefont{J.-M.} \bibnamefont{Triscone}},
  \bibnamefont{and} \bibinfo{author}{\bibfnamefont{M.}~\bibnamefont{Gabay}},
  \bibinfo{journal}{Physical Review B} \textbf{\bibinfo{volume}{86}},
  \bibinfo{pages}{201105} (\bibinfo{year}{2012}),
  \urlprefix\url{http://link.aps.org/doi/10.1103/PhysRevB.86.201105}.

\bibitem[{\citenamefont{Fête}(2014)}]{phdfete_2014}
\bibinfo{author}{\bibfnamefont{A.}~\bibnamefont{Fête}}, Ph.D. thesis,
  \bibinfo{school}{University of Geneva} (\bibinfo{year}{2014}).

\bibitem[{\citenamefont{Anderson}(1958)}]{anderson_absence_1958}
\bibinfo{author}{\bibfnamefont{P.~W.} \bibnamefont{Anderson}},
  \bibinfo{journal}{Physical Review} \textbf{\bibinfo{volume}{109}},
  \bibinfo{pages}{1492} (\bibinfo{year}{1958}), ISSN \bibinfo{issn}{0031-899X},
  \urlprefix\url{https://link.aps.org/doi/10.1103/PhysRev.109.1492}.

\bibitem[{\citenamefont{Abrahams et~al.}(1979)\citenamefont{Abrahams, Anderson,
  Licciardello, and Ramakrishnan}}]{abrahams_scaling_1979}
\bibinfo{author}{\bibfnamefont{E.}~\bibnamefont{Abrahams}},
  \bibinfo{author}{\bibfnamefont{P.~W.} \bibnamefont{Anderson}},
  \bibinfo{author}{\bibfnamefont{D.~C.} \bibnamefont{Licciardello}},
  \bibnamefont{and} \bibinfo{author}{\bibfnamefont{T.~V.}
  \bibnamefont{Ramakrishnan}}, \bibinfo{journal}{Physical Review Letters}
  \textbf{\bibinfo{volume}{42}}, \bibinfo{pages}{673} (\bibinfo{year}{1979}),
  ISSN \bibinfo{issn}{0031-9007},
  \urlprefix\url{https://link.aps.org/doi/10.1103/PhysRevLett.42.673}.

\bibitem[{\citenamefont{Lagendijk et~al.}(2009)\citenamefont{Lagendijk,
  Tiggelen, and Wiersma}}]{lagendijk_fifty_2009}
\bibinfo{author}{\bibfnamefont{A.}~\bibnamefont{Lagendijk}},
  \bibinfo{author}{\bibfnamefont{B.~v.} \bibnamefont{Tiggelen}},
  \bibnamefont{and} \bibinfo{author}{\bibfnamefont{D.~S.}
  \bibnamefont{Wiersma}}, \bibinfo{journal}{Physics Today}
  \textbf{\bibinfo{volume}{62}}, \bibinfo{pages}{24} (\bibinfo{year}{2009}),
  ISSN \bibinfo{issn}{0031-9228},
  \urlprefix\url{https://physicstoday.scitation.org/doi/10.1063/1.3206091}.

\bibitem[{\citenamefont{Giamarchi and Schulz}(1988)}]{giamarchi_anderson_1988}
\bibinfo{author}{\bibfnamefont{T.}~\bibnamefont{Giamarchi}} \bibnamefont{and}
  \bibinfo{author}{\bibfnamefont{H.~J.} \bibnamefont{Schulz}},
  \bibinfo{journal}{Physical Review B} \textbf{\bibinfo{volume}{37}},
  \bibinfo{pages}{325} (\bibinfo{year}{1988}), ISSN \bibinfo{issn}{0163-1829},
  \urlprefix\url{https://link.aps.org/doi/10.1103/PhysRevB.37.325}.

\bibitem[{\citenamefont{Aleiner et~al.}(2002)\citenamefont{Aleiner, Brouwer,
  and Glazman}}]{aleiner_quantum_2002}
\bibinfo{author}{\bibfnamefont{I.~L.} \bibnamefont{Aleiner}},
  \bibinfo{author}{\bibfnamefont{P.~W.} \bibnamefont{Brouwer}},
  \bibnamefont{and} \bibinfo{author}{\bibfnamefont{L.~I.}
  \bibnamefont{Glazman}}, \bibinfo{journal}{Physics Reports}
  \textbf{\bibinfo{volume}{358}}, \bibinfo{pages}{309} (\bibinfo{year}{2002}),
  ISSN \bibinfo{issn}{0370-1573},
  \urlprefix\url{http://www.sciencedirect.com/science/article/pii/S0370157301000631}.

\end{thebibliography}

\end{document}